\documentclass[11pt,a4paper]{iopart}
\usepackage[utf8]{inputenc}
\expandafter\let\csname equation*\endcsname\relax

\expandafter\let\csname endequation*\endcsname\relax 
\usepackage{amsmath}
\usepackage{amsfonts}
\usepackage{amssymb}
\usepackage{makeidx,bm}
\usepackage{graphicx}
\usepackage{iopams}
\usepackage{todonotes}
\usepackage{hyperref}

\usepackage{soul}

\def\bea{\begin{eqnarray}}
\def\eea{\end{eqnarray}}

\def\la{\langle}
\def\ra{\rangle}

\begin{document}
\title{Dynamical fluctuations of a tracer coupled to active and passive particles}

\author{Ion Santra}

\address{Raman Research Institute, Bengaluru 560080, India}

\begin{abstract}
We study the induced dynamics of an inertial tracer particle elastically coupled to passive or active Brownian particles. We integrate out the environment degrees of freedom to obtain the exact effective equation of the tracer---a generalized Langevin equation in both cases. In particular, we find the exact form of the dissipation kernel and effective noise experienced by the tracer and compare it with the phenomenological modeling of active baths used in previous studies. We show that the second fluctuation-dissipation relation (FDR) does not hold at early times for both cases. However, at finite times, the tracer dynamics violate (obeys) the FDR for the active (passive) environment. We calculate the linear response formulas in this regime for both cases and show that the passive medium satisfies an equilibrium fluctuation response relation (FRR), while the active medium does not---we quantify the extent of this violation explicitly. We show that though the active medium generally renders a nonequilibrium description of the tracer, an effective equilibrium picture emerges asymptotically in the small activity limit of the medium. We also calculate the mean squared velocity and mean squared displacement of the tracer and report how they vary with time. 
\end{abstract}

\section{Introduction}
The effective behavior of a system depends crucially on the dynamical nature of its environment. It is well established that if a system is in an equilibrium environment, the linear response of an observable in the presence of small external perturbation is governed by the equilibrium correlations of the same observable. This is popularly called the fluctuation-response relation (also sometimes called the fluctuation-dissipation theorem of the first kind). As a corollary, it is also known that the random force driving the velocity fluctuations of the system is related to the friction (dissipation) experienced by the system. This is known as the fluctuation-dissipation relation (also known in the literature as the fluctuation-dissipation relation of the second kind)~\cite{kubo1966fluctuation,kubo,zwanzigbook,
balakrishnan1979fluctuation,maes2014second}.
%
%
%
Descriptions of such effective motion of a system in an equilibrium environment can be derived starting from the microscopic descriptions of the environment and the system environment coupling and thereafter integrating out the fast environmental degrees of freedom~\cite{zwanzig1973nonlinear,kawasaki1973simple}. Popular and famous ways of doing this include the Feynman-Vernon model~\cite{feynman1963theory} or the Caldeira-Leggett model~\cite{caldeira1983path}, which describe the dynamics of a system (particle) elastically coupled to an environment consisting of a set of non-interacting harmonic oscillators. Integrating out the oscillator degrees of freedom, one can arrive at a generalized Langevin equation for the time evolution of the position of the system. Further, by using the equilibrium correlations of the oscillators, the validity of the fluctuation relations can be explicitly shown.

An interesting question is what happens if the environment is out of equilibrium. The examples of such nonequilibrium environments are widespread in the world of complex systems---a collection of active particles~\cite{collectionactive1,bechinger2016active}, glassy medium~\cite{glass}, sheared fluid~\cite{shearedfluid}, inter-cellular medium~\cite{intracellular} etc. It is known that the familiar forms of the equilibrium relations connecting the fluctuation-dissipation and fluctuation-response of the system do not hold anymore~\cite{fdtviolation1,fdtviolation2,fdtviolation3,
caprini2021fluctuation,caprini2021generalized}. In fact, there are no general forms for the fluctuation relations and they depend on the exact dynamical nature of the environment and how the system is coupled to it~\cite{basu2015statistical}. Thus, providing an effective description of a system in a nonequilibrium environment is a problem highly sought-after. Moreover, transport properties of extended systems connected to active environments have also gained interest recently~\cite{santra2022activity}, and an accurate description of such environments has thus become very important. This is evident from the huge body of works involving experiments in microrheology and biophysics, whose main objective is to understand how the nonequilibrium features of the active environment are reflected on the dynamics of a tracer particle~\cite{collectionactive1,fdtviolation2,soni2003single,maggi2014generalized,
maggi2017memory,turlier2016equilibrium,berner2018oscillating,seyforth2022nonequilibrium,
angelani2011effective,
kasyap2014hydrodynamic,leptos2009dynamics,patteson2016particle,kurihara2017non}. These experiments have, in turn, triggered a number of numerical and theoretical works~\cite{gazuz2009active,desposito2009subdiffusive,morozov2014enhanced,fodor2014energetics,
demery2014generalized,maes2015friction,steffenoni2016interacting,chaki2019effects,
knevzevic2020effective,belan2021active,burkholder2017tracer,
burkholder2019fluctuation,pietzonka2017entropy,kanazawa2020loopy,
granek2020bodies,ye2020active,maes2020fluctuating,ilker2021long,
mousavi2021active,abbaspour2021enhanced,
banerjee2022tracer,granek2022anomalous,abbasi2022non,shea2022passive,
guevara2022brownian}. Importantly, in most of the studies involving fluctuating environments, the effect of the environment is modeled phenomenologically, motivated by experiments, simulations~\cite{fodor2014energetics,knevzevic2020effective,chaki2019effects,
desposito2009subdiffusive,abbasi2022non} and the derivations of the effective motion of a tracer in a nonequilibrium environment starting from the microscopic dynamics and interaction has been limited~\cite{demery2014generalized,maes2020fluctuating,
granek2022anomalous,guevara2022brownian}.
However, such studies provide great insight into what roles the parameters of the fluctuating environment have on the dynamics of the tracer.

In this paper, we investigate the dynamics of a tracer in a fluctuating environment by explicitly integrating the degrees of freedom of the environment. We consider a simple microscopic model where the system, a tracer particle, is connected to a collection of non-self-interacting particles (environment), following their own stochastic dynamics, by harmonic springs. 
In particular, we consider two cases where these stochastic dynamics are Markovian and non-Markovian---modeled by Brownian (passive) and active particles, respectively; each active particle is modeled by a Langevin equation with colored noise, having an exponentially decaying autocorrelation. The harmonic springs render the equations of motion of the tracer and environmental particles to be linear, which helps in integrating out the environmental degrees of freedom exactly to get an effective equation of motion---a generalized Langevin equation for the tracer. We find the exact forms of the effective noise and dissipation experienced by the tracer and show that, though the effective noise correlations contain terms that are not time translation invariant, they decay very fast, and the random force (noise) on the tracer reaches a stationary state. We show that the familiar phenomenological models used to describe active environments in earlier studies are recovered in the strong coupling limit (coupling time-scale much smaller than the active time-scale). Using the exact forms of the dissipation kernel and the stationary noise correlation of the tracer, we find that the traditional forms of the equilibrium FDT are obtained when the environment is made of passive particles. On the other hand, the traditional forms of the FDR are significantly modified for the active particle environment. We compute the linear response of the tracer in the presence of external perturbations explicitly to find that the traditional equilibrium description fits when the environment is passive, while there are stark modifications for the active environment. We calculate all these relations explicitly. Finally, we measure the two most common and easily measurable observables, namely, the mean squared velocity (msv) and the mean squared displacement (msd) of the tracer, and compare the differences in the active and passive environments. In particular, we find that at times shorter than that set by the coupling time-scale, the tracer obeys the usual underdamped behavior.
 However, beyond the coupling time-scale, the tracer follows the motion of the environment---they show a diffusive motion similar to that of the individual environmental particles with the same diffusion constant, albeit with damped oscillations. These oscillations are a result of the visco-elastic nature of the set-up (viscous effects produced by the environmental particles and the elastic effects caused by the coupling springs).
Interestingly, these oscillations vanish for highly active environments (active time-scales much larger than the coupling time-scales). 
Compared to the previous analytical studies, which use a weak coupling limit~\cite{maes2020fluctuating,guevara2022brownian}, adiabatic perturbation theory~\cite{granek2022anomalous} the calculations presented in this paper are exact. Moreover, our derivation of the effective equation of the tracer applies to any choice of the active motion of the environmental particles, unlike the previous works, which take a particular choice of dynamics for the environmental particles.
However, it must be mentioned that though the procedure is exact and simpler compared to the trajectory-based response formalism and adiabatic perturbation theory used in previous works, its applicability beyond harmonic interactions is very difficult.

The paper is organized as follows. We introduce the model and derive the effective generalized Langevin equation for the tracer in Sec.~\ref{s:model}
. In Sec.~\ref{sec:noisecorr}, we compute the autocorrelations of the effective noise appearing in the generalized Langevin equation of the tracer for both active and passive environments. We investigate the linear response formula for the tracer in the presence of small external perturbation and derive the modified response formulas in Sec.~\ref{s:fdt}. In Sec.~\ref{s:msvmsd} we discuss the temporal dependence of mean squared velocity and displacement of the tracer. Finally, we summarize and conclude in Sec.~\ref{s:conc}.

\section{Model}\label{s:model}
Let us consider a particle of mass $m$ coupled elastically to $N$ overdamped particles, which can be either active or passive, via isotropic harmonic springs of spring constant $k_i$. The model is inspired by the typical microrheology experiments which study the motion of a tracer bead in myosin-activated actin networks~\cite{fdtviolation2,turlier2016equilibrium,wirtz2009particle,
lee2010passive,gardel2006prestressed}. The harmonic interaction can also be viewed as an approximation for some other complicated potential $U[x(t)-y_i(t)]$, with a short range, about its minimum. This can be seen by expanding $U(x)$ in a Taylor series about its minimum and keeping terms up to second order; $U''(x_0)$ [$x_0$ denotes the minimum of $U(x)$] serves as an effective spring constant for the medium~\cite{Caprini2020}.
 In this paper, we consider the dynamics along one direction only; however, generalization to higher dimensions is straightforward. The equations of motion of the combined system are given by,
\begin{align}
m\ddot{x}(t)&=-\sum_ik_i\Big(x(t)-y_i(t)\Big)\label{eom:probe}\\
\dot{y}_i(t)&=-\frac{k_i}{\gamma}\Big(y_i(t)-x(t)\Big)+\zeta_i(t)\quad i=1,2\dotsc,N\label{eom:bath1}
\end{align}
where $x$ denotes the tracer displacement while $y_i$ denotes the displacement of the $i$th environment particle; $\gamma$ denotes the damping coefficient of the bath particles; and $\zeta_i(t)$ denotes the random noise of the environemental particles. Note that there is no interaction among the degrees of freedom of the environmental particles.

We consider two cases where the environmental particles are (i) Brownian particles and (ii) active particles. When the environmental particles are Brownian, $\zeta(t)$ is modeled by a delta-correlated noise with zero mean,
\begin{align}
\la\zeta^b_i(t)\zeta^b_j(t')\ra=2D\delta(t-t')\,\delta_{ij}\label{corr:bm}
\end{align}
where the diffusion coefficient $D=K_BT/\gamma$. Active particles, on the other hand, are modeled by stochastic noises which have zero mean and exponentially decaying autocorrelation,
\begin{align}
\la\zeta^a_i(t)\zeta^a_j(t')\ra=v_0^2\,e^{-\alpha|t-t'|}\,\delta_{ij}.\label{corr:ap}
\end{align}
The exponentially decaying autocorrelation  is a hallmark of active particle motion~\cite{BechingerRev,Santra_2022}. Popular models like run-and-tumble particles~\cite{Tailleur2008,Malakar2017,Santra2020} and active Ornstein-Uhlenbeck particles~\cite{Wijland2021} exhibit exponentially decaying correlation of the propulsion velocity at all times. Moreover, the active Brownian particles~\cite{Franosch2016,Basu2018} and direction reversing active Brownian particles~\cite{Santra2020,Santra2021_trap} also follow such similar exponentially decaying noise autocorrelation in the stationary state, see also \ref{app:aparticles}. The characteristic time scale $\alpha^{-1}$ of this decay is a measure of the activity of the particle---smaller $\alpha$ implies larger correlation time, which, in turn, implies higher activity. 
\begin{figure}
\centering\includegraphics[width=0.5\hsize]{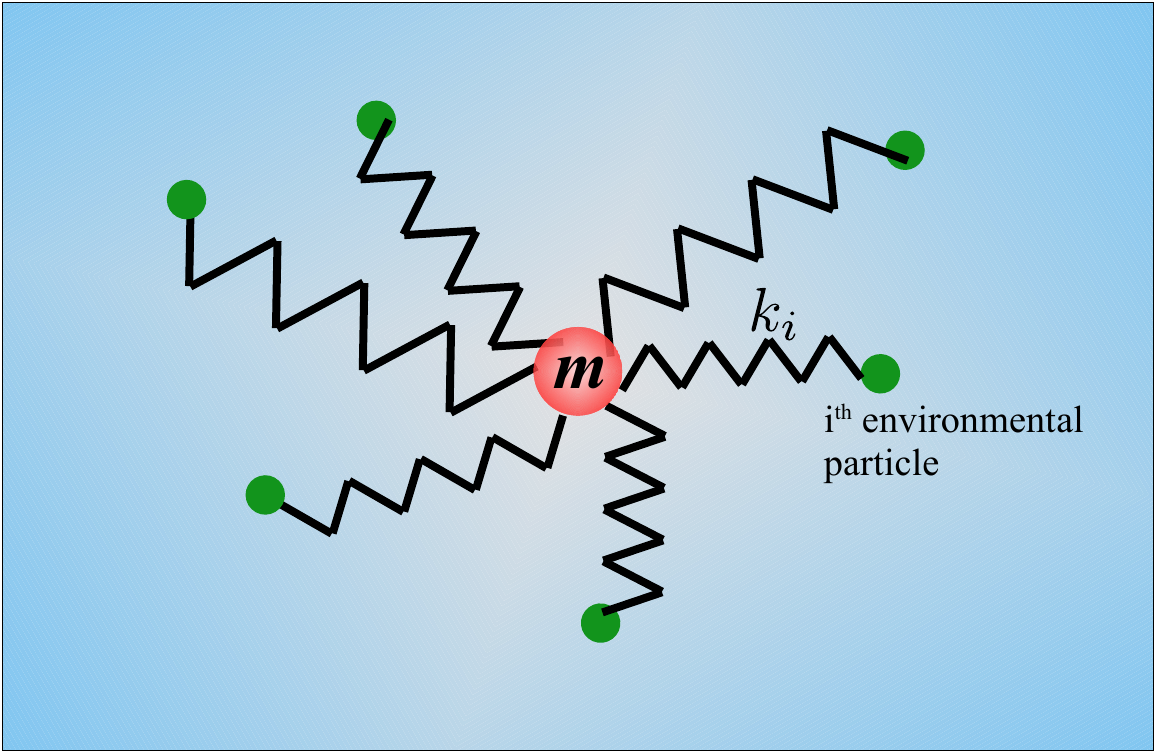}
\caption{A schematic representation of the model: the red shaded circle denotes the tracer of mass $m$ modeled \eref{eom:probe} and the solid green circles denote the environmental particles modeled by \eref{eom:bath1}.}
\end{figure}

In the following, we first derive the equations of motion of the tracer by explicitly integrating out the environmental degrees of freedom $y_i(t)$. To this end, we first note that the equation of motion of the bath particles \eref{eom:bath1} are ordinary first-order linear differential equations with inhomogeneous terms. It has a general solution,
\begin{align}
y_i(t)=y_i(0)e^{-\lambda_i t}+e^{-\lambda_i t}\int_0^t ds\, e^{\lambda_i s}\Big(\lambda_i x(s)+\zeta_i(s) \Big),
\label{eom:bath2}
\end{align}
where $\lambda_i=k_i/\gamma$.
Putting \eref{eom:bath2} in \eref{eom:probe}, we get,
\begin{align}
m\ddot{x}(t)=\sum_i k_i y_i(0) e^{-\lambda t}-\sum_i \Big[k_ix(t)-k_i\lambda_i e^{-\lambda_i t}\int_0^t ds\, e^{\lambda_i s}x(s)-k_i e^{-\lambda_i t}\int_0^t ds\, e^{\lambda_i s}\zeta_i(s)\Big]
\label{eq:autocorr-active}
\end{align}
Thus, the effective dynamics of the tracer depend on the initial condition of the bath particles in the form of a time-dependent velocity that decays exponentially and hence does not affect the long time dynamics of the tracer and we have,
\begin{align}
m\ddot{x}(t)=-\sum_i\,k_ix(t)+\sum_i\Bigg(\frac{k_i^2}{\gamma} e^{-\lambda_i t}\int_0^t ds\, e^{\lambda_i s}x(s)+k_i e^{-\lambda_i t}\int_0^t ds\, e^{\lambda_i s}\zeta_i(s)\Bigg).
\end{align}
Performing an integration by parts on the second integral on the rhs of the above equation, we arrive at a generalized Langevin equation for the tracer dynamics,
\begin{align}
m\ddot{x}(t)=-\int_0^t ds\,\Gamma(t-s)\,\dot{x}(s)+\Omega(t),
\label{eom_effgen}
\end{align} 
where 
\begin{align}\Gamma(\tau)=\sum_i\,k_ie^{-\lambda_i\tau}\label{gamma}
\end{align}
 denotes the memory kernel and 
\begin{align}
\Omega(\tau)= \int_0^t ds\, \sum_i\,k_i\,e^{-\lambda_i (t-s)}\zeta_i(s)
\label{omegadef}
\end{align}
denotes the effective noise experienced by the tracer.
It is important to note that the calculations \eref{eom:bath2} to \eref{eom_effgen} holds for both passive and active environment. Moreover, \eref{eom_effgen},\eref{gamma} and \eref{omegadef} are exact and can be applied to any choice of $\zeta_i(t)$ corresponding to different active dynamics [see \ref{app:aparticles}].
However, the autocorrelation of the effective noise $\Omega(t)$ is different for both cases, which we investigate in the following section. 

\section{Effective noise correlations and second fluctuation-dissipation relation}\label{sec:noisecorr}
In this section, we study the autocorrelations of the effective noise $\Omega(t)$ for both active and passive environments. We need specific forms of $\zeta(t)$ for this. Let us introduce the notation for the two different kinds of environment---physical quantities with the superscript $b$ corresponds to an environment of Brownian particles, while superscript $a$ corresponds to that made of active particles. Using \eref{omegadef} it follows that the effective noise autocorrelation of the tracer is given by the double integral,
\begin{align}
\la \Omega^l(t)\Omega^l(t') \ra=\sum_i k_i^2\,e^{-\lambda_i (t+t')}\int_0^t ds_1 \,\int_0 ^ {t'} ds_2\,e^{\lambda_i (s_1+s_2)}\left\la\zeta^l_i(s_1)\zeta^l_i(s_2) \right\ra\quad\text{where }l=a,\,b.
\label{omegaac}
\end{align}
\subsection{Medium of Brownian particles}
First, we consider the case that the environment consists of Brownian particles. The autocorrelation of the effective noise $\Omega(t)$ in this case can be obtained by using  $\la\zeta^b_i(t)\zeta^b_i(t')\ra=2D\delta(t-t')$ for the Brownian particles in \eref{omegaac}, to get,
\begin{align}
\la \Omega^b(t)\Omega^b(t') \ra&=K_BT \sum_i\,k_i\Big( e^{-\lambda_i|t-t'|}-e^{-\lambda_i(t+t')}\Big)\\
&=K_BT\, \Gamma(|t-t'|)+K_B T\sum_i\,k_i\,e^{-\lambda_i(t+t')}.\label{modifyFDT1}
\end{align}
The presence of the second term on the rhs of the above equation indicates a violation of the standard fluctuation-dissipation theorem at very short times $t,t'\ll \{\lambda_i^{-1}\}$. However, the contribution from the additional term goes to zero as $t,t'>\{\lambda_i^{-1}\}$ (henceforth, we will call this limit as the stationary limit), and we get the equilibrium fluctuation-dissipation relation,
\begin{align}
\la \Omega^b(t)\Omega^b(t') \ra&=K_BT \sum_i\,k_i\, e^{-\lambda_i|t-t'|}=K_BT\, \Gamma(t-t').\label{eqFDT}
\end{align}
\subsection{Medium of active particles}
Next, let us consider the case where the environment is active. The autocorrelation of the effective noise can be obtained by using the noise correlations of the active particles $\la\zeta_i(t)\zeta_i(t')\ra=v_0^2\,e^{-\alpha|t-t'|}$ in \eref{omegaac}. Upon doing the integrals, we finally get (for $\alpha\neq\lambda$),
\begin{align}
\la \Omega^a(t)\Omega^a(t') \ra=\sum_i\frac{v_0^2k_i^2}{\lambda_i^2-\alpha^2}\Bigg[e^{-\alpha|t-t'|}-\frac{\alpha}{\lambda_i}e^{-\lambda_i|t-t'|}&-e^{-(\lambda_i t+\alpha t')}-e^{-(\lambda_i t'+\alpha t)}\cr
&+\frac{(\alpha-\lambda_i)}{\lambda_i}e^{-\lambda_i(t+t')}  \Bigg].
\label{omegaA_corr1}
\end{align}
Unlike the previous case of the passive environment, where the stationary limit of the effective noise autocorrelation satisfies the equilibrium FDT, the above expression clearly indicates a violation of FDT, even in the stationary limit. In this limit, the noise correlations reduce to,
\begin{align}
\la \Omega^a(t)\Omega^a(t') \ra=\sum_i\frac{v_0^2k_i^2}{\alpha^2-\lambda_i^2}\left[\frac{\alpha}{\lambda_i}e^{-\lambda_i|t-t'|}-e^{-\alpha|t-t'|}\right].
\label{omegaA_corr2}
\end{align}
This is the modified Fluctuation dissipation relation followed by a tracer elastically coupled to an active environment. For $\alpha=\lambda$, one needs to do the integrals in \eref{omegaac} starting with $\alpha=\lambda_i=\lambda$. The stationary noise correlation, in that case, is given by,
\begin{align}
N^{-1}\la \Omega^a(t)\Omega^a(t') \ra=\frac{v_0^2k^2}{2\lambda}|t-t'|\, e^{-\lambda|t-t'|}
\end{align}

\noindent It is important to note two limits of \eref{omegaA_corr2}:
\begin{itemize}
\item Large spring constant ($\lambda=k/\gamma\gg\alpha$): In this limit the noise correlation \eref{omegaA_corr2} reduces to,
\begin{align}
N^{-1}\la \Omega^a(t)\Omega^a(t') \ra=v_0^2\gamma^2 e^{-\alpha|t-t'|}.
\end{align}
Additionally, in this limit, the memory kernel \eref{gamma} also reduces to a delta function,
\begin{align}
N^{-1}\Gamma(\tau)\approx\gamma \delta(\tau),
\end{align}
which leads to the effective equation motion of the tracer to be,
\begin{align}
M\ddot{x}(t)=-\gamma\dot{x}(t)+\bar\Omega^a(t).\label{activetracer}
\end{align}
where $M=m/N$ and $\bar\Omega^a(t)$ is an exponentially correlated colored noise, having the same correlation time as the constituent active particles of the environment. The form of~\eref{activetracer} has been used in previous works~\cite{maggi2014generalized,
chaki2019effects,santra2022activity}, the emergence of the same from a microscopic model of active environment justifies these phenomenological models.
Moreover, \eref{activetracer} is familiar with the equation of standard underdamped active particles. 
Usually, a problem for studying single active particle trajectories is that the active particles are small in size and have a finite lifetime. The result, that the motion of a tracer particle in an environment of active particles emulates the motion of those active particles, might be of great help to the experimentalists studying single active particle trajectories.

\item Passive limit: This limit corresponds to $\alpha\to\infty$, $v_0\to\infty$, such that $D_{\text{eff}}=v_0^2/(\alpha)$ is finite. Putting these limits in the noise correlation \eref{omegaA_corr2}, we obtain an effective passive limit of the active environment, 
\begin{align}
\la \Omega^a(t)\Omega^a(t') \ra&=\frac{v_0^2}{\alpha}\sum_i\,k_i e^{-\lambda_i|t-t'|}\cr
&=K_BT_{\text{eff}}\, \Gamma(t-t').
\label{modFDTactive}
\end{align}
The above equation in the stationary state reduces to the well-known equilibrium form of the FDT, with the effective temperature being given by, $D_{\text{eff}}=K_B T_{\text{eff}}/\gamma $.
\end{itemize}
Note that, since the environmental particles are treated effectively as independent, it is sufficient to consider the coupled dynamics of only one active or passive particle with the tracer and in the following sections, we will consider the effect of a single passive or active particle on the tracer.

\section{Linear Fluctuation-Response formulas}\label{s:fdt}
Linear response theory describes the leading order behavior of a system in the presence of some external perturbations. The Fluctuation-Response relations (FRRs) relate the response of a system observable to external perturbations with the fluctuations of the same observable in the unperturbed system. In this section, we investigate how the tracer in the active or passive environment reacts to external perturbations by calculating the linear response formulas for its position and velocity and deriving the FRRs explicitly.

 Let us consider an external force $f(t)$ acting on the tracer. Thus, the effective equation of motion of the tracer is now given by,
\begin{align}
m\ddot{x}(t)=-\int_0^t ds\Gamma(t-s)\dot{x}(s)+\Omega(t)+f(t),
\label{eom_force}
\end{align}
where the memory kernel and effective noises are the same as defined before. Now, the velocity and displacement response functions are defined by the change in the average velocity and displacement due to this external perturbation $f(t)$,
\begin{align}
\Delta v(t)&=\la v(t)\ra_f-\la v(t)\ra_0 =\int_{0}^tR^v(t-t')f(t'),\\ 
\Delta x(t)& =\la x(t)\ra_f-\la x(t)\ra_0=\int_{0}^tR^x(t-t')f(t').
\end{align}
Here the subscripts $f$ and $0$ denote the averages in the presence and absence of the perturbation, respectively. The above equations have a much simpler form in the Fourier space,
\begin{align}
\Delta \tilde v(\omega)=\tilde{R}^v(\omega)\tilde{f}(\omega),\quad \text{and}\quad
\Delta \tilde x(\omega)=\tilde{R}^x(\omega)\tilde{f}(\omega),
\label{resp1}
\end{align}
where the Fourier transform is defined by, $h(t)=\int_{-\infty}^{\infty}e^{-i\omega t}\tilde{h}(\omega)/(2\pi)$ and $\tilde{h}(\omega)=\int_{-\infty}^{\infty}e^{i\omega t}h(t)$.

Alternatively, $\Delta \tilde v(\omega)$ and $\Delta\tilde{x}(\omega)$ can be obtained directly from the effective Langevin equations of the probe \eref{eom_force}. To this end, we take a Fourier transform of the probe equation of motion to get,
\begin{align}
\tilde{v}(\omega)=\frac{\tilde{\Omega}(\omega)+\tilde{f}(\omega)}{\tilde\Gamma(\omega)-i \omega m},\quad\text{and}\quad
\tilde{x}(\omega)=\frac{\tilde{\Omega}(\omega)+\tilde{f}(\omega)}{-m\omega^2-i\omega\tilde\Gamma(\omega)}.\label{ft_v}
\end{align}
Note that the absence of the external force is realized by simply putting $\tilde{f}(\omega)=0$ in the above equation. Using \eqref{ft_v} and the fact that $\la \Omega(t)\ra=0$, we get,
\begin{align}
\Delta \tilde v(\omega)=\frac{\tilde{f}(\omega)}{\tilde\Gamma(\omega)-i \omega m},\quad\text{and}\quad
\Delta \tilde x(\omega)=\frac{\tilde{f}(\omega)}{-m\omega^2-i\omega\tilde\Gamma(\omega)}.\label{resp2}
\end{align}
Further, comparing \eqref{resp1} and \eqref{resp2}, we get the velocity and position response in Fourier space as,
\begin{align}
\tilde{R}^v(\omega)&=\frac{1}{\tilde\Gamma(\omega)-i \omega m},\label{respv}\\
\tilde{R}^x(\omega)&=\frac{1}{-m\omega^2-i\omega\tilde\Gamma(\omega)}.\label{respx}
\end{align}
Thus the linear response function is independent of what kind of bath the probe is connected to.

Now, let us compute the position and velocity autocorrelations in the absence of external perturbations in Fourier space. Using equations \eref{ft_v} along with $\tilde{f}(\omega)=0$, we can formally write the autocorrelations as,
\begin{align}
\la \tilde v^i(\omega)\tilde v^i(\omega')\ra_0&=\frac{\la \tilde\Omega^i(w)\tilde\Omega^i(w')\ra}{(\tilde\Gamma(\omega)-i\omega m)(\tilde\Gamma(\omega')-i\omega' m)},\label{vauto}\\
\la \tilde x^i(\omega)\tilde x^i(\omega')\ra_0&=\frac{\la \tilde\Omega^i(w)\tilde\Omega^i(w')\ra}{(-m\omega^2-i\omega\tilde\Gamma(\omega))(-m\omega'^2-i\omega'\tilde\Gamma(\omega'))},\label{xauto}
\end{align}
where the superscript $i=b,\,a$ depends on whether the environment is passive or active. Thus, we need the autocorrelations of the effective noises $\Omega^{i}$ in the Fourier space. These can be easily obtained by Fourier transforms of  \eqref{eqFDT} and \eqref{modFDTactive} given by,
\begin{align}
\la\tilde{\Omega}^i(\omega)\tilde{\Omega}^i(\omega')  \ra &=2\pi C^i(\omega)\delta(\omega+\omega')\quad\text{where }i=a,b \label{cc}\\
\text{with    }
C^b(\omega) &=K_B T\frac{2k\lambda}{\lambda^2+\omega^2}\label{cb}\\
\text{and    }
C^a(\omega) &=\frac{2\alpha\,v_0^2 k^2}{\lambda^2-\alpha^2}\left( \frac{1}{\alpha^2+\omega^2}-\frac{1}{\lambda^2+\omega^2}\right).\label{ca}
\end{align}

\subsection{Passive environment}
The velocity autocorrelation in Fourier space, using \eref{vauto} and \eref{cc}, is given by,
\begin{align}
\la \tilde v^b(\omega)\tilde v^b(\omega')\ra_0&=\frac{2\pi C^b(\omega)~\delta(\omega+\omega')}{(\tilde\Gamma(\omega)-i\omega m)(\tilde\Gamma(-\omega)+i\omega m)}.
\end{align}
Further, using the Fourier transform of the dissipation kernel $\tilde\Gamma(t)$ [given in \eref{gamma}],  $\tilde\Gamma(\omega)=k/(\lambda-i\omega)$, and \eref{cb} we find,
\begin{align}
\la \tilde v^b(\omega)\tilde v^b(\omega')\ra_0=4\pi K_BT\delta(\omega+\omega')\frac{k^2/\gamma}{(k-m\omega^2)^2+m^2\omega^2\lambda^2}.
\end{align}
It is straightforward to show using the form of $\tilde R^v(\omega)$ in \eref{respv} that, 
\begin{align}
\la \tilde v^b(\omega)\tilde v^b(\omega')\ra_0&=4\pi K_BT\delta(\omega+\omega')\Re \left[ \tilde R^v(\omega)\right]\label{FRR_bv},
\end{align}
where $\Re[\dotsb]$ denotes the real part of a complex number.

Again, from \eref{ft_v} and \eref{cc}, the position autocorrelation in Fourier space is given by,
\begin{align}
\la \tilde x^b(\omega)\tilde x^b(\omega')\ra_0&=\frac{2\pi C^b(\omega)~\delta(\omega+\omega')}{(m\omega^2+i\omega\tilde\Gamma(\omega))(m\omega^2-i\omega\tilde\Gamma(-\omega))}.
\end{align}
Using the explicit form of $C_b$ and $\tilde{\Gamma}(\omega)$, it can be shown [following steps exactly the same as the velocity autocorrelation discussed above] that,
\begin{align}
\la \tilde x^b(\omega)\tilde x^b(\omega')\ra_0&=4\pi K_B T\delta(\omega+\omega')\frac{k^2/\gamma}{\omega^2((k-m\omega^2)^2+m^2\omega^2\lambda^2)}\cr
&=\frac{\pi K_B T\delta(\omega+\omega')}{\omega}\Im \left[ \tilde R^x(\omega)\right],\label{FRR_bx}
\end{align}
where $\Im[\dotsb]$ denotes the imaginary part of a complex number.

Equations of the structure of \eref{FRR_bv} and \eref{FRR_bx} relate the velocity and position response of the tracer in a passive environment in the presence of external perturbations to velocity and position fluctuations in the absence of the external perturbations, are called FRRs. It is important to note that both the FRRs obtained here are in the familiar form, as seen in equilibrium systems.

\subsection{Active environment}
The velocity autocorrelation in Fourier space using \eref{vauto} and \eref{cc} is given by,
\begin{align}
\la \tilde v^a(\omega)\tilde v^a(\omega')\ra_0&=\frac{2\pi C^a(\omega)~\delta(\omega+\omega')}{(\tilde\Gamma(\omega)-i\omega m)(\tilde\Gamma(\omega')-i\omega' m)}
\label{act_velcorr}
\end{align}
Upon using the explicit forms of $C^a(\omega)$ and $\tilde\Gamma(\omega)$, we get,
\begin{align}
\la \tilde v^a(\omega)\tilde v^a(\omega') \ra&=\frac{4\pi\,\alpha v_0^2\delta(\omega+\omega')}{\alpha^2-\lambda^2} \frac{1}{m^2\omega^2\lambda^2+(k-m\omega^2)^2}\left(1-\frac{\lambda^2+\omega^2}{\alpha^2+\omega^2}  \right)\cr
&=\frac{4\pi\,\alpha v_0^2\delta(\omega+\omega')}{\alpha^2-\lambda^2}\frac{\gamma}{k^2}\Re\left[\tilde R^v(\omega)\right]\left(1-\frac{\lambda^2+\omega^2}{\alpha^2+\omega^2}  \right)
\label{act_vac}
\end{align}

Similarly, the position autocorrelation, given by,
\begin{align}
\la \tilde x^a(\omega)\tilde x^a(\omega')\ra_0&=\frac{2\pi C^a(\omega)~\delta(\omega+\omega')}{(m\omega^2+i\omega\tilde\Gamma(\omega))(m\omega^2-i\omega\tilde\Gamma(-\omega))}.
\end{align}
can be evaluated using the expressions of $C^a(\omega)$ and $\tilde\Gamma(\omega)$. We finally have,
\begin{align}
\la \tilde x^a(\omega)\tilde x^a(\omega') \ra&=\frac{4\pi\,\alpha v_0^2\delta(\omega+\omega')}{\alpha^2-\lambda^2}\frac{\gamma}{\omega k^2}\Im\left[\tilde R^x(\omega)\right]\left(1-\frac{\lambda^2+\omega^2}{\alpha^2+\omega^2}  \right)
\label{act_xac}
\end{align}

Equations \eref{act_vac} and \eref{act_xac} are clearly different from the well-known equilibrium forms of the Fluctuation-Response relations. This is expected since the environment, consisting of active particles, is itself out of equilibrium. In fact, \eref{act_vac} and \eref{act_xac} denote the modified FRRs of a tracer in an active environment. It is perhaps useful to rewrite the modified FRRs as,
\begin{align}
\la \tilde v^a(\omega)\tilde v^a(\omega') \ra&= 4\pi K_B \hat{T}\delta(\omega+\omega')\Re\left[\tilde R^v(\omega)\right]\left(1-\frac{\lambda^2+\omega^2}{\alpha^2+\omega^2}  \right),\label{frrv}\\
\la \tilde v^a(\omega)\tilde v^a(\omega') \ra&= \frac{4\pi K_B \hat{T}}{\omega}\delta(\omega+\omega')\Im\left[\tilde R^v(\omega)\right]\left(1-\frac{\lambda^2+\omega^2}{\alpha^2+\omega^2}  \right),\label{frrx}
\end{align}
where $\hat{T}=\frac{\alpha v_0^2\gamma/k^2}{\alpha^2-\lambda^2}$. Clearly, the second term in the above equations denotes the extent of violation of the well-known fluctuation-response relations. It must be mentioned that these explicit quantitative measures of FRR violations are related to the energy dissipation rate in nonequilibrium stochastic systems~\cite{haradasasa} and can be used to study the thermodynamics of such tracer dynamics in active environments in the future.  

We end this section with a discussion on the effective passive limit, $\alpha\to\infty$, $v_0\to\infty$, such that $D_{\text{eff}}=v_0^2/(\alpha)=K_BT_{\text{eff}}/\gamma$ is finite, discussed earlier at the end of Sec.~\ref{sec:noisecorr}. In this limit, the FRR violation terms in \eref{frrv} and \eref{frrx} go to zero and $\hat{T}\to T_{\text{eff}}$, and we get back the equilibrium-like fluctuation-response relations with the effective temperature $T_{\text{eff}}$.
\begin{align}
\la \tilde v^b(\omega)\tilde v^b(\omega')\ra_0&=4\pi K_BT_{\text{eff}}\delta(\omega+\omega')\Re \left[ \tilde R^v(\omega)\right],\\
\la \tilde x^b(\omega)\tilde x^b(\omega')\ra_0&=\frac{4\pi K_BT_{\text{eff}}}{\omega}\delta(\omega+\omega')\Im \left[ \tilde R^x(\omega)\right].
\end{align}

\section{Mean squared velocity and displacement}\label{s:msvmsd}
In this section, we investigate the two most common experimentally relevant observables, namely, the mean squared velocity (msv),
\begin{align}
\la(\Delta_v)^2\ra=\left\la\left(v(t)-v(t_0)\right)^2\right\ra=\left(\la v(t)^2\ra+\la v(t_0)^2\ra-2\la v(t)v(t_0)\ra\right),
\end{align}
and the mean squared displacement (msd),
\begin{align}
\la(\Delta_x)^2\ra=\left\la\left(x(t)-x(t_0)\right)^2\right\ra=\left(\la x(t)^2\ra+\la x(t_0)^2\ra-2\la x(t)x(t_0)\ra\right).
\end{align}

The velocity autocorrelation in real-time can be obtained by doing an inverse Fourier transform of $\la \tilde v^i(\omega)\tilde v(\omega')\ra$ [where $i=b,a$ denoting passive and active environments, respectively],
\begin{align}
\la v^i(t)v^i(t_0) \ra=&\frac{1}{4\pi^2}\int_{-\infty}^{\infty}d\omega \int_{-\infty}^{\infty}d\omega' e^{-i(\omega t+\omega' t_0)}\la \tilde v^i(\omega)\tilde v(\omega')\ra\cr
=&\frac{1}{2\pi}\int_{-\infty}^{\infty}d\omega  e^{-i\omega(t-t_0)}\frac{C^i(\omega)}{(\tilde\Gamma(\omega)-i\omega m)(\tilde\Gamma(-\omega)+i\omega m)},\label{vtvt'},
\end{align}
where in getting the last line, we have used \eref{vauto} and \eref{cc}, followed by performing the delta function integration over $\omega'$.
Similarly, the position autocorrelation in real-time, using \eref{xauto} and \eref{cc} can be written as,
\begin{align}
\la x^i(t)x^i(t_0) \ra=&\frac{1}{2\pi}\int_{-\infty}^{\infty}d\omega  e^{-i\omega(t-t_0)}\frac{C^i(\omega)}{(m\omega^2+i\omega\tilde\Gamma(\omega))(m\omega^2-i\omega\tilde\Gamma(-\omega))}.\label{xtxt'}
\end{align}
Now, since from the above equations, it is clear that the two-point temporal correlations of position and velocity depend only on the time difference, we can put $t_0=0$, for simplicity, during the calculation of the msv and the msd. This, along with the symmetry of the integrands in \eref{vtvt'} and \eref{xtxt'} leads to,
\begin{align}
\la(\Delta^i_v)^2\ra=\frac{2}{\pi}\int_{0}^{\infty}d\omega\left(1-\cos(\omega t)\right)\frac{(\lambda^2+\omega^2)C^i(\omega)}{(m\omega^2-k)^2+m^2\omega^2\lambda^2}\label{msvgen}\\
\la(\Delta^i_x)^2\ra=\frac{2}{\pi}\int_{0}^{\infty}\frac{d\omega}{\omega^2}\left(1-\cos(\omega t)\right)\frac{(\lambda^2+\omega^2)C^i(\omega)}{(m\omega^2-k)^2+m^2\omega^2\lambda^2}\label{msdgen}
\end{align}
 The above integrals are hard to evaluate exactly to get closed-form expressions; however, in the following, we numerically evaluate and compare them with numerical simulations performed with \eref{eom:probe} and \eref{eom:bath1}. For numerical simulations, we model the environment by a one-dimensional run-and-tumble particle [see \ref{app:sim} for more details]; however, since the msd and msv depend only on the two-point correlations of the active noise, the results will hold for other models of active dynamics like active Ornstein-Uhlenbeck particles and active Brownian particles which have exponentially decaying two-point correlations of the active noise. We also find the asymptotic forms of the leading order behavior of the msv and msd when the time-scales are well separated.

 \subsection{Passive environment}

\begin{figure}
\includegraphics[width=\hsize]{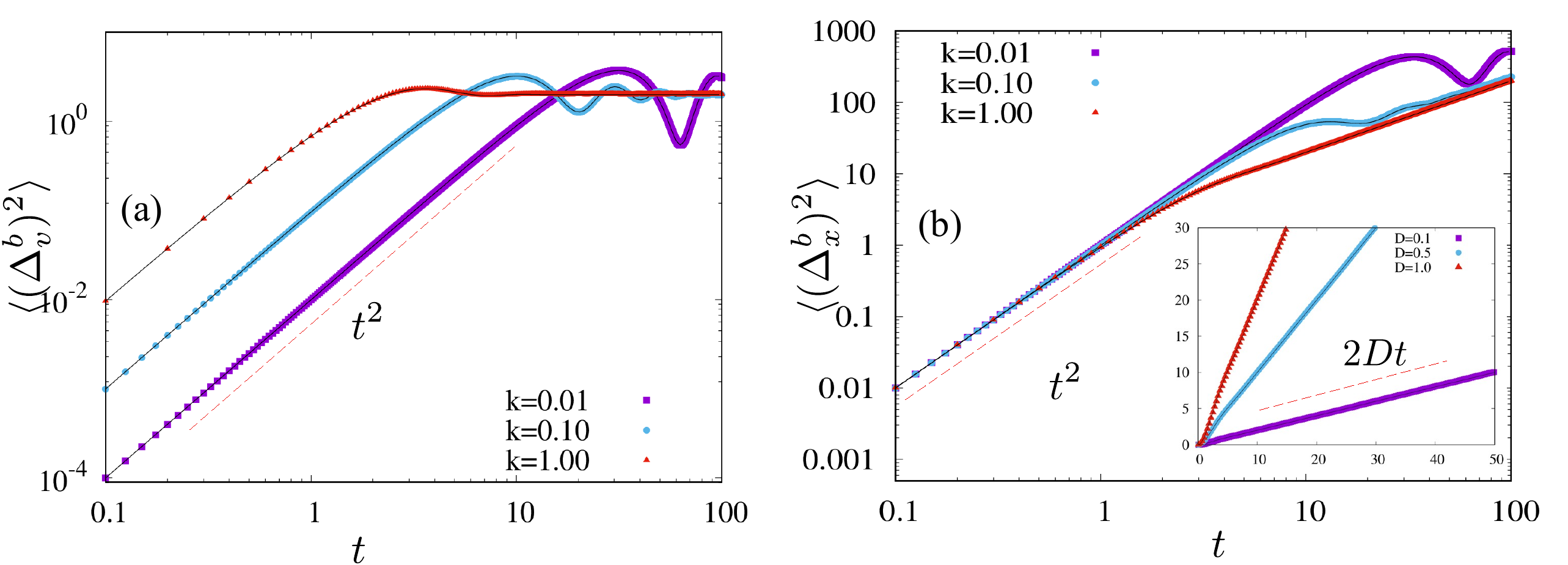}
\caption{Plots for the msv and msd of the tracer in the passive environment consisting of one Brownian particle coupled to a tracer particle of unit mass. (a) shows shows the mean squared velocity for $D=1$, $\gamma=1$ and three different values of $k$. The symbols correspond to the numerical simulations, while the solid black lines indicate the analytical prediction \eref{pass:msv}. (b) shows the mean squared displacement for $D=1$, $\gamma=1$ and three different values of $k$. The symbols correspond to the numerical simulations, while the solid black lines indicate the analytical prediction \eref{pass:msv}. The inset highlights the long-time diffusive behavior; the dashed lines correspond to the predicted analytical form in \eref{pass_msdL}.}\label{f:bm}
\end{figure}

Let us consider the passive environment first. Using the form of $C_b(\omega)$ (given in \eref{cb}) in \eref{msvgen}, we have,
\begin{align}
\la(\Delta^b_v)^2\ra=\frac{4Dk^2}{\pi}\int_{0}^{\infty}d\omega\left(1-\cos(\omega t)\right)\frac{1}{(m\omega^2-k)^2+m^2\omega^2\lambda^2},\label{pass:msv}
\end{align}
where $D=K_BT/\gamma$ is the diffusion constant of the environment particles as defined before. As discussed previously, it is difficult to perform the above integral exactly. However, it can be evaluated numerically; this is compared with the numerical simulations performed on the full system [tracer + environment, Eqs.~\eref{eom:probe} and \eref{eom:bath1}]  [see~\ref{app:sim} for details on the numerical simulations] in \fref{f:bm} (a) and shows excellent agreement. The following observations are made along with some asymptotic results (the asymptotic calculations and arguments are provided in \ref{app:calc}).

At short times $t\ll \lambda^{-1}$ the msv grows as,
\begin{align}
\la(\Delta^b_v)^2\ra\approx At^2, \quad\text{with }A=\frac{2Dk^2}{\pi}\int_{0}^{\infty}d\omega\frac{w^2}{(m\omega^2-k)^2+m^2\omega^2\lambda^2}.
\label{pass_msvS}
\end{align}
This $t^2$ growth is indicated in \fref{f:bm} (a) by red dashed lines. At long times $t\gg \lambda^{-1}$, the msv exhibits a damped oscillatory behavior and saturates to a constant value,
\begin{align}
\la(\Delta^b_v)^2\ra=v_{ss}^2=\frac{4Dk^2}{\pi}\int_{0}^{\infty}d\omega\frac{1}{(m\omega^2-k)^2+m^2\omega^2\lambda^2}.\label{pass_msvL}
\end{align}
Interestingly the constant value $v_{ss}^2$ is independent of the coupling strength $k$. 

Similarly, the mean squared displacement of the tracer in the passive environment can be obtained by using \eref{cb} in \eref{msdgen},
\begin{align}
\la(\Delta^b_x)^2\ra=\frac{4Dk^2}{\pi}\int_{0}^{\infty}\frac{d\omega}{\omega^2}\left(1-\cos(\omega t)\right)\frac{1}{(m\omega^2-k)^2+m^2\omega^2\lambda^2}.
\end{align}
mathThe integral on the rhs of the above equation is difficult to evaluate exactly but can be evaluated numerically. The comparison of the numerical integration with the simulations of the full system \eref{eom:probe} and \eref{eom:bath1} is shown in \fref{f:bm} (b) and shows very good agreement. We make the following observations. The msd also grows ballistically at shorter times $t\ll \lambda^{-1}$,
\begin{align}
\la(\Delta^b_x)^2\ra\approx \frac{v_{ss,b}^2}2 \,t^2,\label{pass_msdS}
\end{align}
where $v_{ss,b}^2$ is the stationary value of msv at long times, given in \eref{pass_msvL}. This ballistic behavior at short times (indicated by dashed lines in \fref{f:bm} (b) [main plot] by red dashed lines.) crosses over to a diffusive behavior,
\begin{align}
\la(\Delta^b_x)^2\ra\approx 2Dt+8D\left(2\frac mk-\frac{m^2}{\gamma^2}\right)\label{pass_msdL}
\end{align}
at large times ($t\gg \lambda^{-1}$). This behavior is illustrated in the inset of \fref{f:bm} (b) by red dashed lines. 
 \subsection{Active environment}
\begin{figure}
\includegraphics[width=\hsize]{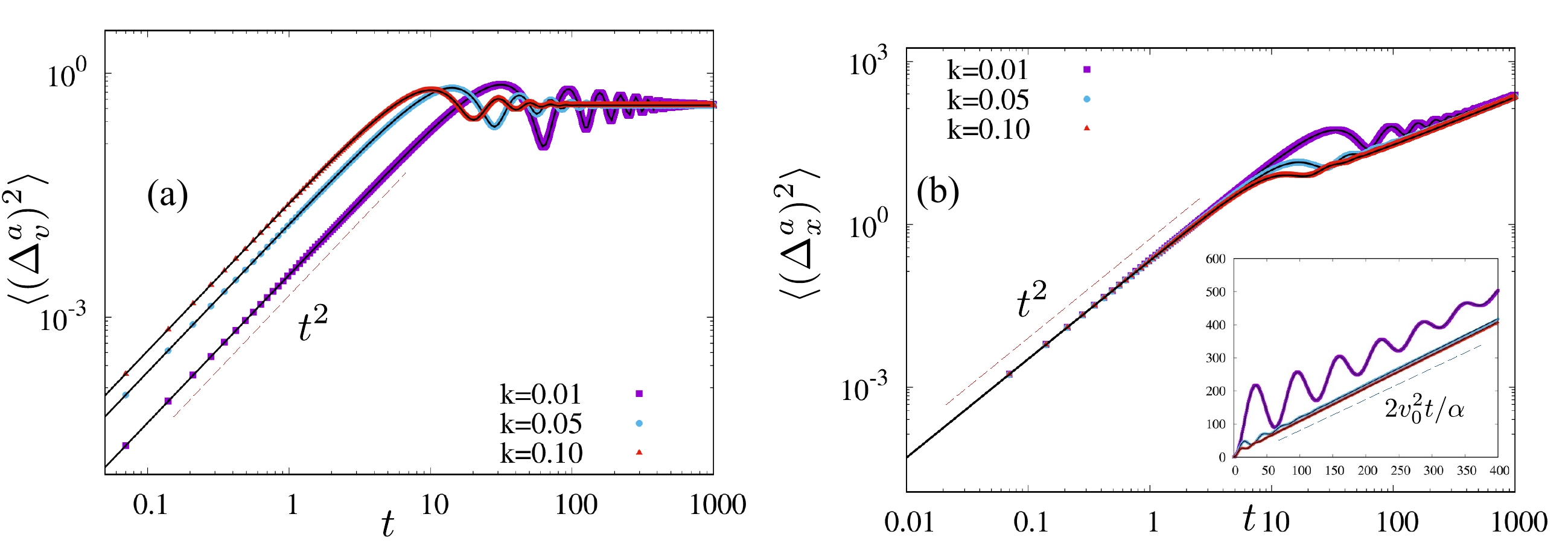}
\caption{Plots showing the mean squared velocity and mean squared displacement of the tracer in the active environment coupled to a run-and-tumble particle for $m/k\gg\alpha^{-1}$. (a) shows the msv for 
$\gamma=1,\,\alpha=1$ and three different values of $k$. The symbols correspond to the numerical simulations, while the solid black lines indicate the analytical prediction \eref{act:msv}. (b) shows the msd $\gamma=1,\,\alpha=1$ and three different values of $k$. The symbols correspond to the numerical simulations, while the solid black lines indicate the analytical prediction \eref{msd:act}. The inset shows the same data in linear scale, with the dashed lines indicating the asymptotic behavior predicted in \eref{msd_a_lt}.}
\label{f:act1}
\end{figure}

\begin{figure}
\centering\includegraphics[width=\hsize]{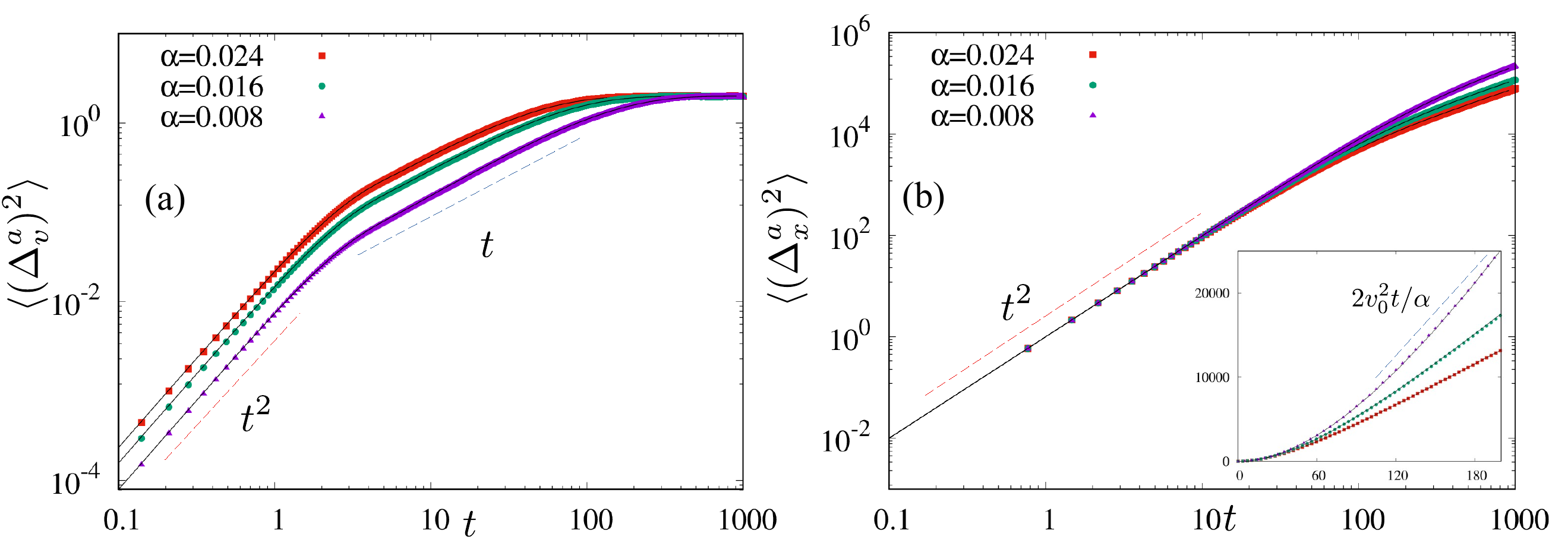}
\caption{Plots showing the mean squared velocity and mean squared displacement of the tracer in the active environment coupled to a run-and-tumble particle for $\lambda^{-1}\ll\alpha^{-1}$. (a) shows the msv for 
$\gamma=1,\,k=1$ and three different values of $\alpha$. The symbols correspond to the numerical simulations, while the solid black lines indicate the analytical prediction \eref{act:msv}. The presence of three temporal regimes with different behavior, as predicted in \eref{msvak} can be seen clearly; the $\sim t^2$ and $\sim t$ behavior in the short and intermediate time regimes are indicated by dashed lines. (b) shows the msd $\gamma=1,\,k=1$ and three different values of $\alpha$. The symbols correspond to the numerical simulations, while the solid black lines indicate the analytical prediction \eref{msd:act}. The inset shows the same data in linear scale, with the dashed lines indicating the asymptotic behavior predicted in \eref{msd_a_lt}.}\label{f:act2}
\end{figure}

 In active environment, using the expression of $C^a(\omega)$ in \eref{ca} in \eref{msvgen}, the msv can be written as,
 \begin{align}
 \la(\Delta^a_v)^2\ra=\frac{4\alpha v_0^2k^2}{\pi(\alpha^2-\lambda^2)}\int_{0}^{\infty}d\omega\left(1-\cos(\omega t)\right)\frac{\left(1-\frac{\lambda^2+\omega^2}{\alpha^2+\omega^2}\right)}{(m\omega^2-k)^2+m^2\omega^2\lambda^2}\label{act:msv}
 \end{align}
This integral cannot be performed exactly; however, a comparison of the numerical evaluation of the integral with simulations (where the active particles are modeled by one-dimensional run-and-tumble particles) in \fref{f:act1} (a) and \fref{f:act2} (a) shows excellent agreement.

In the case of active particles, in addition to the time-scale set by the coupling parameter $\lambda^{-1}$, the active time-scale of the medium $\alpha^{-1}$ sets up another time-scale. The msv has different behavior depending on which time-scale is larger. If the coupling time-scale is larger than the active time scale, i.e., $\alpha>\lambda^{-1}$, 
\begin{align}
\la(\Delta^a_v)^2\ra= C t^2 \quad\text{for }t\ll \lambda^{-1}
\label{msvka1}
\end{align}  
where
\begin{align}
C=\frac{2\alpha v_0^2k^2}{\pi(\alpha^2-\lambda^2)}\int_{0}^{\infty}d\omega\frac{\left(1-\frac{\lambda^2+\omega^2}{\alpha^2+\omega^2}\right)}{(m\omega^2-k)^2+m^2\omega^2\lambda^2}.\label{msvact:C}
\end{align}
 At times larger than the coupling time scale $\lambda^{-1}$, the msv exhibits a damped oscillatory behavior and saturates to the constant value,
\begin{align}
\la(\Delta^a_v)^2\ra\approx v^2_{ss,a}=\frac{4\alpha v_0^2k^2}{\pi(\alpha^2-\lambda^2)}\int_{0}^{\infty}d\omega\frac{\left(1-\frac{\lambda^2+\omega^2}{\alpha^2+\omega^2}\right)}{(m\omega^2-k)^2+m^2\omega^2\lambda^2}
\label{msvka2}
\end{align}  
The theoretical prediction \eref{act:msv} for $k<\alpha$ is shown in \fref{f:act1}(a) along with numerical simulations.

On the other hand, if the active time-scale $\alpha^{-1}$ is larger than the coupling time-scale, i.e., $\alpha<\lambda$, then the three regimes show three distinct behaviors,
\bea
\la(\Delta^a_v)^2\ra\approx\left \{ \begin{split}
&Ct^2\quad t\ll \lambda^{-1}\\
&\alpha v_0^2 t \quad \lambda^{-1}\ll t \ll \alpha^{-1}\\
&v^2_{ss,a} \quad \alpha^{-1}\ll t
\end{split}\right.
\label{msvak}
\eea
where $C$ is again given in \eref{msvact:C}. A comparison of numerical simulations and our theoretical prediction \eref{act:msv} for $k<\alpha$ is shown in \fref{f:act2}(a). The behaviors in the different regimes as predicted by \eref{msvak} are indicated by dashed lines in the same figure.

The msd of the tracer in the active environment, using \eref{ca} and \eref{msdgen}, is given by,
\begin{align}
\la(\Delta^a_x)^2\ra=\frac{4\alpha v_0^2k^2}{\pi(\alpha^2-\lambda^2)}\int_{0}^{\infty}\frac{d\omega}{\omega^2}\left(1-\cos(\omega t)\right)\frac{\left(1-\frac{\lambda^2+\omega^2}{\alpha^2+\omega^2}\right)}{(m\omega^2-k)^2+m^2\omega^2\lambda^2}\label{msd:act}
\end{align} 
This is compared with numerical simulations in \fref{f:act1}(b) and \fref{f:act2}(b) and shows excellent agreement.

Unlike msv, the msd does not show different temporal behavior in the three different time regimes. The short and intermediate time regimes show the same dynamical behavior of msd irrespective of the values of the coupling and activity,
\begin{align}
\la(\Delta^a_x)^2\ra= \frac{v^2_{ss,a}}{2}\, t^2 \quad \text{for }  t\ll \text{max}\,[\alpha^{-1},\lambda^{-1}]\label{msd_a_st}
\end{align}
This ballistic growth is shown in \fref{f:act1}(b) and \fref{f:act2}(b) $\lambda^{-1}\gg\alpha^{-1}$ and $\lambda^{-1}\ll\alpha^{-1}$, respectively.

At larger times $t\gg \text{max}\,[\alpha^{-1},\lambda^{-1}]$ the msd exhibits a diffusive behavior,
\begin{align}
\la(\Delta^a_x)^2\ra= 2D_\text{{eff}}\,t+O(1),\label{msd_a_lt}
\end{align}
where $D_\text{{eff}}=v_0^2/\alpha$. 
This prediction is shown in \fref{f:act1}(b) and \fref{f:act2}(b) $\lambda^{-1}\gg\alpha^{-1}$ and $\lambda^{-1}\ll\alpha^{-1}$, respectively.

Interestingly, depending on the values of the coupling and active time-scales a damped oscillatory behavior is seen in the msd of the tracer. If the coupling time scale is large, then a damped oscillatory behavior about the diffusive behavior \eref{msd_a_lt} is observed in the long-time regime $t\gtrsim \lambda^{-1}$ [see inset of \fref{f:act1}]. The damped oscillatory behavior vanishes completely, when the active time-scale $\alpha^{-1}$ is larger, and the tracer exhibits the bare diffusive motion \fref{f:act2}.

\subsection{Discussion}
Let us try to understand the above analysis qualitatively by comparing the two cases. For the tracer in a passive environment, there is only one characteristic time scale of the system, which is set by the coupling $\lambda^{-1}=\gamma/k$. For this reason, at time scales much smaller than $\lambda^{-1}$, the msv and the msd follow the usual behavior of an underdamped particle with memory. At time scales larger than the $\lambda^{-1}$, the visco-elastic (viscous dissipation caused by the dynamics of the environment particles and elastic effects caused by the connecting springs) effects of the environment come into play, which leads to the damped oscillatory behavior. The overall behavior of the tracer in this regime is, however, a reflection of the motion of the environment---reaching a constant value for msv and diffusive for msd. In fact, the diffusion constant at large times is also the same as that of the environmental particles.

For the active environment, on the other hand, there are two time-scales---one is the same coupling time-scale $\lambda^{-1}$, while the other is the active time-scale of the environmental particles $\alpha^{-1}$. It is important to note that active particles have an effective diffusive behavior at times much larger than their corresponding active time-scales. So, if the coupling time-scale is much larger than the active time-scale $\lambda^{-1}\gg\alpha^{-1}$, then the activity of the environment does not have any effect on the effective behavior of the tracer, and the msv and msd exhibit behaviors similar to that seen in a passive environment. However, if the active time-scale is larger than the coupling time-scale, then the extreme short-time behavior $t\ll \lambda^{-1}$ is the same as the previous case. At times larger than the coupling time-scale a non-trivial intermediate regime is seen. The msv grows linearly with time. Interestingly, the damped oscillatory behavior, seen in all the previous cases for $t>\lambda^{-1}$, is largely suppressed when $\lambda\ll\alpha$. In fact, they are completely absent in the msd, which shows a ballistic growth $\sim t^2$. This implies that the activity of the environment suppresses the viscoelastic effects. The large-time behavior $t\gg \alpha^{-1}$ is again the same: the tracer follows the motion of the environmental active particles $\la (\Delta_x^a)^2\ra=2D_\text{eff} t$, where $D_\text{eff}$ is the long-time effective diffusion coefficients of the individual active particles.

\section{Summary and Conclusion}\label{s:conc}

We study the effective behavior of a tracer particle in passive and active viscoelastic environments. These environments are modeled by a collection of non-interacting overdamped Brownian and active particles, respectively. The tracer is coupled to the environmental particles by harmonic springs. We start from the microscopic equations of motion of the tracer and the environmental particles and integrate the degrees of freedom of the latter to get the effective time dynamics of the tracer. We find that the effective dynamics of the tracer follow a generalized Langevin equation with an exponential memory function and a non-trivial effective noise. We explicitly calculate the effective noise and find the generalized fluctuation-dissipation relations(FDRs). In fact, we show that in the stationary state (times greater than the time-scales of the coupling and environment), the effective tracer dynamics follow the equilibrium FDR in the case of the passive environment, while the standard form is modified for the active environment. However, we find an effective passive limit where the modification terms tend to zero, and we asymptotically get the equilibrium-like FDRs, with an effective temperature.
We calculate the Fluctuation Response Relations, which relate the linear response of an observable of the system to external perturbations in terms of the correlations of the same observables in the absence of perturbations. We find equilibrium-like FRR in the passive environment, while the modified FRR in the active environment is in the form of \cite{haradasasa}. Finally, we look at two observables: the mean squared velocity (msv) and mean squared displacement (msd) of the tracer. We find that, in general, at times shorter than the coupling time-scales, the behavior in both the environments is characterized by the usual underdamped behavior, while at times larger than the coupling time-scales $\lambda^{-1}=\gamma/k$, the tracer follows the dynamics of the environmental particles. The viscoelastic effect of the medium shows up as damped oscillations about the leading order behavior in this time regime. An exception to the above behavior occurs in the active environment when the active 
time-scale $\alpha^{-1}$ are much larger than the coupling time scale: then in the intermediate regime $\lambda^{-1}<t<\alpha^{-1}$ the msv shows a non-trivial temporal growth while the damped oscillatory behavior vanishes completely.

The analysis carried out here is exact; the simple model of the bilinear coupling between a tracer and active/passive environments renders the exact analytical treatment doable and helps in understanding the effective dynamics of the tracer. In fact, the effective generalized Langevin equation of the tracer obtained here reduces to the various phenomenological modeling of the baths used to study the thermodynamic and transport properties of active reservoirs. Moreover, the exact results obtained in this paper can be used for any dynamical model of active particles.
 the analytical form of the modified fluctuation response relations of the tracer in the presence of the active environment can be used to study the energetics of active environments using the famous Harada-Sasa equality. A very obvious extension of this simple set-up is to introduce interaction among the environmental degrees of freedom and study how the effective tracer dynamics are affected. Another possible extension is the study of anharmonic active medium perturbatively, in the spirit of~\cite{bhadra2016system}. In this work, the activity of the environmental particles has been modeled only by the exponentially decaying noise autocorrelation of the environmental particles. The chosen form of the autocorrelations is the same for different kinds of active dynamics, like run-and-tumble processes, active Ornstein-Uhlenbeck particles, and active Brownian particles. It is thus interesting to study the higher-order dynamical fluctuations to understand how the signature of specific forms of active motion of the environmental particles affects the tracer dynamics. It will also be interesting to study the transport properties of extended systems connected to active environments using the generalized Langevin equation obtained here, in the spirit of \cite{santra2022activity}.
   
\section{Acknowledgements}
The authors thank Urna Basu, Sanjib Sabhapandit, Sayantan Majumdar, and Christina Kurzthaler for useful discussions, comments, and suggestions. 
   
   \appendix
   
  \section{Discussion on microscopic dynamics of active particles}
   \label{app:aparticles}
   
   In this appendix, we introduce the basic active particle models that exhibit exponentially decaying autocorrelation of the form \eref{eq:autocorr-active} in the long-time limit.
   
\begin{itemize}

\item {\bf Run-and-tumble process (RTP)}: 
In this model the particle moves with a constant speed, interrupted by intermittent tumbling events, changing the direction of the velocity randomly~\cite{Santra2020}. The run durations are usually taken from an exponential distribution $\alpha'\,e^{-\alpha't}$. 
In one-dimension~\cite{Tailleur2008,Malakar2017}, this is modelled by a dichotomous noise $\zeta(t)$ in \eref{eom:bath1},
\bea
\zeta(t) = v \sigma(t),
\eea
where $\sigma(t)$ alternates between $1$ and $-1$ with rate $\alpha'$. In this case, $\zeta=\pm v_0$ can take only two discrete values and the corresponding propagator is given by \cite{Santra2021},
\bea
P(\zeta,t|\zeta',0)=\frac{1}{2}\left( 1+\zeta \zeta' e^{-2 \alpha' t}\right).\label{rtp_prop}
\eea  
Clearly, in the stationary state, the two values of $\zeta$ occur with equal probability $1/2$. It is straightforward to see that this process leads to the two point auto-correlation of the form \eref{eq:autocorr-active} with,
\begin{align}
\alpha=2\alpha' \quad\text{and }v_0=v.
\end{align}

\item{\bf Active Ornstein-Uhlenbeck Process (AOUP)}: In this model, the noise $\zeta(t)$ undergoes an Ornstein-Uhlenbeck~\cite{wang1945theory,caprini2022parental},
\bea
\dot \zeta(t) = -\frac 1 {\tau} \zeta + \sqrt{\frac{2D}{\tau^2}}\, \bar\eta(t), \label{eq:aoup}
\eea
where $\bar\eta(t)$ is a Gaussian white noise with $\la \bar\eta(t) \ra =0$ and  $\la \bar\eta(t) \bar\eta(t') \ra =\delta(t-t')$; the diffusion constant $D$ denotes the strength of the noise.  The linear nature of the process and the Gaussian nature of the noise leads to a Gaussian propagator for the noise $\zeta(t)$,
{\bea
{\cal P}(\zeta,t|\zeta',0) = \frac{\exp{\left( -\frac{\tau_j}{2 D}\frac{(\zeta-\zeta' e^{-t/\tau})^2}{1-e^{-2t/\tau}} \right)}}{\Big[2 \pi D(1-e^{-2t/\tau})\Big]^{1/2}}.\label{aoup_prop}
\eea}
Evidently, the noise $\zeta(t)$ has a correlation of the form of \eref{eq:autocorr-active} with,
\begin{align}
\alpha=1/\tau \quad\text{and }v_0 = \sqrt{D/ \tau}. \label{eq:aj_aoup}
\end{align}

\item{\bf Active Brownian process (ABP)}: Here the particle self-propels with a constant speed along a direction $\hat{n}=(\cos\theta,\sin\theta)$, where the internal orientation $\theta$ of the active particle undergoes a Brownian motion~\cite{Basu2018}. The marginal one-dimensional process along $x$ direction is described by the Langevin equation \eref{eom:bath1} with,
\bea
\zeta(t) = v\cos \theta_j(t),\quad\text{with}\quad \dot \theta(t) = \sqrt{2 D^R}\, \bar{\eta}(t).
\eea
Here $\bar{\eta}$ refers to a Gaussian white noise with $\la \bar{\eta}(t) \ra =0$ and  $\la \bar{\eta}(t) \bar{\eta}(t') \ra =\delta(t-t')$. Clearly, $\theta(t)$ undergoes a standard Brownian motion which leads to a Gaussian propagator,
\bea
\mathcal{P}(\theta,t|\theta',0)=\frac{1}{\sqrt{4 \pi D^R t}}\exp{\left(-\frac{(\theta-\theta')^2}{4 D^R t}\right)}.
\eea
The corresponding distribution for $\zeta(t)$ eventually reaches a stationary state with autocorrelation of the form \eref{eq:autocorr-active} with 
\bea
\alpha=D^R,\quad\text{ and } v_0=v/\sqrt{2}.\label{eq:aj_abp}
\eea

\item{ {\bf Direction reversing active Brownian particle (DRABP):}} In this case, the particle does an ABP-like motion, while intermittently reversing its direction completely. The duration between two consecutive reversals is chosen from an exponential distribution $\alpha' e^{-\alpha' t}$. The marginal one-dimensional process along $x$-axis is described by the Langevin equation \eref{eom:bath1} with,
\bea
\zeta(t) = v\,\sigma(t)\cos \theta_j(t),\quad\text{with}\quad \dot \theta(t) = \sqrt{2 D^R}\, \bar{\eta}(t),
\eea
where $\sigma(t)$ is again a dichotomous process which changes stochastically between $\pm 1$ at rate $\alpha'$.
The joint propagator in this case is given by \cite{Santra2021},
\begin{align}
\mathcal{P}(\theta,\sigma,t|\theta',\sigma'0)=\frac{1}{2\sqrt{4 \pi D^R t}}\exp{\left(-\frac{(\theta-\theta')^2}{4 D^R t}\right)}\left( 1+\sigma \sigma' e^{-2 \alpha' t}\right)
\end{align}
The corresponding distribution for $\zeta(t)$ eventually reaches a stationary state with autocorrelation of the form \eref{eq:autocorr-active} with 
\bea
\alpha=(D^R+2\alpha'),\quad\text{ and } v_0=v/\sqrt{2}.\label{eq:aj_abp}
\eea

\end{itemize}

   \section{Asymptotic forms of msv and msd}\label{app:calc}
In this section we give the details of the computation of the asymptotic forms of msv and msd given in the main text. 
\subsection{Short time behavior}
Let us investigate the short time behavior first. To this end, let us consider an integral of the form,
\begin{align}
I={\cal A}\int_0^{\infty}d\omega\, (1-\cos(\omega t)){\cal G}(\omega),\label{gen:form}
\end{align}
 The above integral can always be written as,
  \begin{align}
  I={\cal A}\int_0^{\infty}d\omega\, \sum_{k=1}^{\infty}(-1)^{k+1}\frac{(\omega t)^{2k}}{(2k)!}{\cal G}(\omega).
  \end{align}
Now, at short times, this can be well approximated by,
 \begin{align}
  I=\frac{{\cal A}t^2}{2}\int_0^{\infty}d\omega\, \omega^2{\cal G}(\omega).\label{shorttime:general}
  \end{align} 
 Both the msd and msv have the common general form as in \eref{gen:form} with different ${\cal G}(\omega)$. Since $I\sim t^2$ at short times irrespective of the form of ${\cal G}(\omega)$, the short-time behavior for msd and msv always grows as $\sim t^2$. The coefficients can be easily calculated using \eref{shorttime:general} and the particular forms of ${\cal G}(\omega)$.

\subsection{Behavior at late times} 
 At very large times ($t$ is greater than all other time-scales), the major contribution to the integral comes from small values of $|\omega|\to 0$.
The asymptotic behavior in this regime depends on the form of ${\cal G}(\omega)$. In fact, the msv and msd show very different behaviors. Let us consider two integrals,
 \begin{align}
 {\cal J}_v={\cal A}\int_0^{\infty}d\omega\, (1-\cos(\omega t))g(\omega),\label{msv:latetime}\\
 {\cal J}_x={\cal A}\int_0^{\infty}d\omega\, (1-\cos(\omega t))\frac{g(\omega)}{\omega^2}.\label{msd:latetime}
 \end{align}
The equations \eref{pass:msv} and \eref{act:msv} are in same form as  
 \eref{msv:latetime} with,
 \begin{align}
 g(\omega)=\begin{cases}
 \left[(m\omega^2-k)^2+m^2\omega^2\lambda^2\right]^{-1}\quad &\text{for passive environment}\\
\left(1-\frac{\lambda^2+\omega^2}{\alpha^2+\omega^2}\right)\left[(m\omega^2-k)^2+m^2\omega^2\lambda^2\right]^{-1}\quad &\text{for active environment}
 \end{cases}
\end{align}  
and
\bea
 {\cal A}=\left\{\begin{split}
 \frac{4Dk^2}\pi\quad &\text{for passive environment}
 \cr
\frac{4\alpha v_0^2 k^2}{\pi(\alpha^2-\lambda^2)}\quad &\text{for active environment}\\
 \end{split}\right.
\eea 
  In both these cases, the poles of $g(w)$ are either general complex numbers of the form $a+ib$, or purely imaginary of the form $ib$. The contribution to the integral $\int_0^{\infty}d\omega\, \cos(\omega t)g(\omega)$ always decays to zero exponentially with time. Thus, to find the leading order behavior of msv at large-times, we can drop the term proportional to $\cos(\omega t)$ to obtain the stationary values of msv,
\begin{align}
v^2_{ss,b}&\approx \frac{4Dk^2}\pi\int_0^{\infty}d\omega\,\left[(m\omega^2-k)^2+m^2\omega^2\lambda^2\right]^{-1},\\
v^2_{ss,a}&\approx \frac{4\alpha v_0^2 k^2}{\pi(\alpha^2-\lambda^2)}\pi\int_0^{\infty}d\omega\,\left(1-\frac{\lambda^2+\omega^2}{\alpha^2+\omega^2}\right)\left[(m\omega^2-k)^2+m^2\omega^2\lambda^2\right]^{-1},
\end{align}
for tracer in passive and active environment, respectively.

For msd \eref{msd:latetime}, on the other hand, there is always a second order pole at $\omega=0$, and thus the leading order contribution at large times, comes from the integral,
\begin{align}
{\cal J}_x\approx{\cal A}\,g(0)\int_0^{\infty}d\omega\, \frac{1-\cos(\omega t)}{\omega^2}=\frac{{\cal A}\,g(0)\pi}{2}\,t
\end{align}
This is the long time diffusive behavior reported in the main text. The explicit forms for passive and active environments can be obtained by using appropriate values of ${\cal A}$ and $g(\omega)$.

\subsection{Intermediate regime for active environment}
The msv of the tracer in active environment is given by \eref{act:msv}, which can be rewritten as,
\begin{align}
\la(\Delta^a_v)^2\ra=\frac{4\alpha v_0^2k^2}{\pi}\int_{0}^{\infty}d\omega\left(1-\cos(\omega t)\right)\frac{\left(\frac{1}{\alpha^2+\omega^2}\right)}{(m\omega^2-k)^2+m^2\omega^2\lambda^2}.
\end{align}
The intermediate regime is characterized by $\lambda^{-1}\ll t\ll\alpha^{-1}$. Thus, for msv we can approximate the integral by,
\begin{align}
\la(\Delta^a_v)^2\ra=\frac{4\alpha v_0^2k^2}{\pi}\int_{0}^{\infty}d\omega\frac{1-\cos(\omega t)}{\omega^2}\frac{1}{(m\omega^2-k)^2+m^2\omega^2\lambda^2}.
\end{align}
This is again of the form \eref{msd:latetime} and we get the leading order behavior,
\begin{align}
\la(\Delta^a_v)^2\ra=\frac{4\alpha v_0^2}{\pi}\int_{0}^{\infty}d\omega\frac{1-\cos(\omega t)}{\omega^2}=2\alpha v_0^2\, t.
\end{align}
   
   For the mean-squared displacement, which can also be rewritten from \eref{msd:act},
\begin{align}
\la(\Delta^a_v)^2\ra=\frac{4\alpha v_0^2k^2}{\pi}\int_{0}^{\infty}d\omega\frac{1-\cos(\omega t)}{\omega^2}\frac{\left(\frac{1}{\alpha^2+\omega^2}\right)}{(m\omega^2-k)^2+m^2\omega^2\lambda^2}.
\end{align}   
   In this case, though $\alpha$ is very small, $(\alpha^2+\omega^2)$ cannot be approximated by $\omega^2$, like the previous case, because, it changes the order of the pole at $\omega=0$. Thus, the leading order contribution comes from the integral,
   \begin{align}
   \la(\Delta^a_v)^2\ra=\frac{4v_0^2}{\pi\,\alpha}\int_{0}^{\infty}d\omega\frac{1-\cos(\omega t)}{\omega^2}=\frac{ 2v_0^2}{\alpha}\, t=2D_{\text{eff}}t.
   \end{align}
   The next subleading order term of the msd in this regime comes from the pole $\omega=i\alpha$, which is purely imaginary and leads to an exponentially decaying contribution without any oscillations. This accounts for the absence of oscillations in msd in this regime.

\section{Details of numerical simulation}\label{app:sim}
For numerical simulations, the passive environment are modeled by Brownian particles, whose position increment $\Delta y$ in a time step $\Delta t$ is given by,
\begin{align}
\Delta y=\sqrt{2 D \Delta t}\,\eta_t-k(y-x)\Delta t,
\end{align}
where $\eta_t$ is a random number drawn from a Gaussian distribution of zero mean and unit variance.
For the active particle environment, we model the environmental particles as run-and-tumble particles; the active noise in \eref{eom:bath1} is modeled by a dichotomous noise $\zeta(t)$ that switches between $+v_0$ and $-v_0$ at rate $\alpha_{\text{RT}}$. The autocorrelation in this case is given by,
\begin{align}
\la \zeta(t)\zeta(t') \ra=v_0^2\exp\left(-2\alpha_{\text{RT}}|t-t'|\right).
\end{align}
The increment $\Delta y$ in a time step $\Delta t$ in this case is,
\begin{align}
\Delta y=\zeta\,\Delta t-k(y-x)\Delta t, \quad\text{where }\zeta=\pm 1.
\end{align}
For all the simulations, we used $\Delta t=0.0005$ and sample over $10^5$ trajectories. It is worth pointing out that the initial position of the ocillators do not affect the behavior of the msv and msd in the `stationary' state (of the effective noise of the tracer), observed here.

\section*{References}
\bibliography{ref}   
\end{document}